\documentclass[sn-mathphys]{sn-jnl}

\jyear{2022}%

\theoremstyle{thmstyleone}%
\theoremstyle{thmstyletwo}%

\theoremstyle{thmstylethree}%

\begin{document}

\title[Article Title]{Studies on quantum turbulence with Vinen}


\author*[1,2]{\fnm{Makoto} \sur{Tsubota}}\email{tsubota@omu.ac.jp}



\affil*[1]{\orgdiv{Department of Physics}, \orgname{Osaka Metropolitan University}, \orgaddress{\street{Sugimoto 3-3-138, Sumiyoshiku}, \city{Osaka}, \postcode{558-8585}, \state{Osaka}, \country{Japan}}}




\abstract{In my research career in the field of quantum turbulence, I have been encouraged by Vinen. 
In this article, I review my works motivated by him and my joint collaborations with him. 
Vinen encouraged me in studies on quantum turbulence at zero temperature, Kolmogorov spectrum of a vortex tangle without mutual friction, and 
fully coupled dynamics of quantized vortices and normal fluid.  
Joint works include studies on diffusion of a vortex tangle, Kelvin wave cascade, quantum turbulence created by an oscillating object, and coupled dynamics of tracer particles and quantized vortices. }

\keywords{Quantum turbulence, Quantized vortex, superfluid, two-fluid model}



\maketitle

\section{Introduction}\label{sec1}
Joe Vinen (1930-2022) was a great scientist who made a significant contribution to the field of quantum hydrodynamics and turbulence.
He pioneered this field in its early stages through seminal works on superfluid $^4$He discovering mutual friction  \cite{hallvinen1,hallvinen2}, thermal counterflow \cite{tc1,tc2,tc3,tc4}, and quantized circulation \cite{circulation}. 
After the 1990s, he conducted new studies on quantum turbulence worldwide. 
He has influenced and encouraged many scientists, and I am one of them.
If I had not met him, I would not have studied quantum turbulence.
In this article, I review the works triggered by Vinen and studied in collaboration with him, and discuss how these works led to present studies on quantum turbulence.

I met Vinen for the first time at Shin-Osaka Station in Shinkansen, Japan in October 1998. 
He was invited to Japan by Kimitoshi Kono. 
Following the suggestion of Kono, Vinen visited me in Osaka, Japan.
Before I saw Vinen, I had published several papers on numerical studies of the dynamics of quantized vortices, but I was not familiar with quantum turbulence and was isolated from this field’s global community.
As soon as Vinen met me, he asked me, "Do you know what happens to a vortex tangle at very low temperatures? It decays, contrary to expectations."
Then, Vinen expounded on the latest experimental work on superfluid $^4$He by McClintock {\it et al.} \cite{davis}.
In this study, the authors reported that vortices created by a vibrating grid decayed even below $T \sim $70 mK where mutual friction was not possible.
I was surprised to learn about this experimental work; it 
encouraged me to study quantum turbulence.

The purpose of this article is not to review the studies on quantum turbulence.
For that, we ask the readers to refer to the excellent review articles by Vinen \cite{Vinen02,Vinen10}.
I describe how my work was encouraged by Vinen and our joint work together. 
Of note, all of the numerical calculations described in this article were performed using the vortex filament model (VFM) \cite{JLTPreview}.

\section{Works encouraged by Vinen}\label{sec3}
In the 1990s, several important experimental papers were published on the energy spectrum of quantum turbulence \cite{Maurer98,Stalp99}.
It is well known that the energy spectrum follows the Kolmogorov -5/3 law in a three-dimensional fully developed classical turbulence \cite{Frisch,davidson}.
However, what happens to the energy spectrum of quantum turbulence is not trivial.
Vinen discussed the classical analog of quantum turbulence based on these experiments \cite{Vinen00}.
The key argument was that on length scales significantly greater than both the spacing between the quantized vortices and the scale on which viscous dissipation occurs in the normal fluid, the two fluids were likely to be coupled together and behave like a conventional fluid. The quantized effects of vortices did not appear explicitly on these scales.
On smaller length scales, dissipation had to be considered because the viscosity in the normal fluid caused mutual friction between quantized vortices and the normal fluid, resulting in sound radiation from quantized vortices. 
In their study, Vinen stressed the importance of experiments at very low temperatures because the effects of the normal fluid complicate the problem at finite temperatures.
Vinen also showed a sketch of the energy cascade at low temperatures. On scales larger than the inter-vortex spacing, the energy spectrum demonstrated the conventional Kolmogorov law. The energy was transferred by the cascade of quantized vortices.
On scales smaller than the inter-vortex spacing, the energy was relayed to a cascade of Kelvin waves.  
This concept is widely accepted and has become a standard guiding principle in this field.

Discussions on these topics with Vinen encouraged me to study quantum turbulence numerically, without mutual friction. 
In those days, there was little information about this issue.

\subsection{Quantum turbulence at zero temperature}\label{subsec2}
We performed numerical simulation of vortex tangles without mutual friction \cite{tsubota00}.
The important finding was the characteristic configuration of vortices. 
When mutual friction worked at a finite temperature, the vortex lines became relatively smooth because the mutual friction dissipated its small-scale structure.
Without mutual friction, however, the vortex lines tended to kink. 
Through vortex reconnections, kinked structures were created locally near the reconnection points and excited the Kelvin waves.
Through these processes, the vortex tangle decayed, even without mutual friction.
In the numerical simulation of the VFM, we usually eliminated small vortices whose size became comparable to the numerical space resolution, namely, the distance between neighboring points along the filament. 
Such small vortices continued to be created through the cascade process.
This numerical simulation supports the results presented by Vinen \cite{Vinen00}. 

This work has led to a vast number of theoretical and experimental studies in both superfluid helium and atomic Bose-Einstein condensates \cite{Bagnato}.
The important one of them is the study of types of quantum turbulence. 
Walmsley {\it et al.} studied experimentally dissipation of quantum turbulence in the zero temperature limit \cite{Walmsley07} and led to classification of  two types of quantum turbulence \cite{Walmsley08}, namely ultraquantum turbulence and quasiclassical turbulence, depending on the correlations in the orientation of quantized vortices.
This classification has developed our understanding of quantum turbulence \cite{GolovPNAS}.
 
\subsection{Kolmogorov spectrum of a vortex tangle without mutual friction}\label{subsubsec2}
Subsequently, we numerically studied the energy spectrum of a vortex tangle without mutual friction \cite{Araki02}.
The energy spectrum of quantum turbulence had already been studied by the Gross--Pitaevskii model by Nore {\it et al.} \cite{Nore}, who made a vortex tangle from the Taylor--Green vortex and observed that the spectrum of the incompressible kinetic energy followed the Kolmogorov law in the process of turbulence decay. 
We performed a similar calculation using the VFM and confirmed that the energy spectrum followed Kolmogorov’s law at scales larger than the inter-vortex spacing.
This study was strongly encouraged by Vinen.

In Ref. \cite{Araki02}, we also found numerically that the vortex length distribution in the tangle showed a power-law. 
The purpose of this study was to connect the self-similarity in wavenumber space, namely the energy spectrum, and the self-similarity in the real space, for example, the vortex length distribution.
Our first draft of this paper included a theoretical argument relating the energy spectrum and vortex length distribution, as well as the numerical results.
However, Vinen pointed out the shortcomings of the theoretical argument; thus, we removed it and showed only the numerical results.

\subsection{Fully-coupled dynamics of quantized vortices and the normal fluid}
Since the discovery of thermal counterflow, two turbulence states have been observed when the aspect ratio of the cross-section of the channel was low: T-1 and T-2 states \cite{tough}. 
The T-1 state was thought to be associated with turbulence only in the superfluid, whereas in the T-2 state, both fluids were likely turbulent\footnote{Melotte and
Barenghi performed a linear stability analysis of the normal fluid in the T-1 state and suggested that the laminar normal fluid could become unstable owing to mutual friction \cite{Melotte}.}.
To investigate this mystery, we had to address the fully coupled dynamics of the superfluid and normal fluid by considering the realistic boundary conditions of channels. 
However, the early numerical studies of VFM, namely the full Biot--Savart simulation \cite{Adachi} as well as the simulation based on the localized induction approximation \cite{Schwarz88}, addressed the vortex dynamics under the prescribed normal fluid profile.
There were two reasons why the fully coupled dynamics of the two fluids was not studied in early works. 
First, the numerical scheme of the fully coupled dynamics was difficult.
Second, there was insufficient information on normal fluid profile in T-1 and T-2 states; this lack of information did not motivate theorists to challenge fully coupled dynamics of the two fluids. 

However, visualization experiments in the field of superfluid $^4$He have been conducted and changed the situation \cite{WeiGuo}.  
Vinen was very aware of the importance of visualization experiments  \cite{Vinen10}.
The first visualization of quantized vortices in terms of micron-sized hydrogen particles was performed by Bewley {\it et al.} \cite{Bewley}.
Guo then started visualization experiments on counterflow using metastable helium molecules with Vinen \cite{Guo10}.
What impressed me about this among a series of studies was the observation of the deformation of the velocity profile of the laminar normal fluid, namely, the tail-flattened flow where the tail parts close to the channel boundaries became flattened \cite{Marakov}.  

This observation of the tail-flattened flow prompted me to study the fully coupled dynamics of quantized vortices and normal fluid in the counterflow.
Before the counterflow visualization experiments, there were only a few numerical studies on fully coupled dynamics \cite{KIvotides00,KIvotides07}, which were not relevant to counterflow.
Then, we numerically studied the fully coupled dynamics of the counterflow in a square channel and found that the velocity profile of the normal fluid was deformed significantly, similar to the observation by Marakov {\it et al.} \cite{Marakov},  by a vortex tangle as it became dense  \cite{Yui18}.
However, the spatial grid for the normal fluid flow was rough and was criticized by Vinen.
Guo performed a series of visualization experiments on counterflow using solidified deuterium particles as tracers \cite{Mastracci18,Mastracci19,Mastracci19-2,Tang20}.
In particular, they reported velocity fluctuations of the laminar normal fluid and showed that streamwise fluctuations exhibited higher intensity and longer-range autocorrelation than transverse fluctuations \cite{Mastracci19-2}.
They proposed a simple model of anisotropic fluctuations.
Motivated by this observation, we performed a numerical simulation of the fully coupled dynamics by making the spatial grid of the normal fluid finer so that it could capture the fluctuation properly \cite{Yui20}.
The obtained results for the anisotropic fluctuation of the normal fluid were consistent with the observations and supported the analysis of the simple model.

These recent visualization experiments and numerical simulations furthered the understanding of turbulence in counterflow.
In the near future, we will be able to reveal the issues of the T-1 and T-2 states.
It would be interesting to show the developments to Vinen.

\section{Joint works with Vinen}\label{sec4}
This section describes my collaboration with Vinen.

\subsection{Diffusion of a vortex tangle}
Vinen was interested in the diffusion of an inhomogeneous vortex tangle in its early stages.
This motivation was derived from experiments on quantum turbulence-created vibrating structures \cite{VinenSkrbek}.
A vibrating structure  created a localized vortex tangle near it.
Any interpretation of this experiment requires an understanding of the behavior of inhomogeneous tangles.
Therefore, we numerically studied the diffusion of a vortex tangle \cite{Diffusion}.

A vortex tangle was created under counterflow at 1.6 K in a cube.
Periodic boundary conditions were applied at the faces normal to the flow, while the other surfaces were considered solid.
After a vortex tangle developed, the counterflow was turned off, the temperature was reduced to zero, vortices with parts in half of the cube were removed, and the evolution of the remaining vortices was followed. 
This evolution involved decay and diffusion to the empty side.

Then, we compared the numerical results with the \lq\lq inhomogeneous" Vinen equation.
The original Vinen equation was proposed as the equation of motion for the vortex line density, $L(t)$, of a homogeneous vortex tangle \cite{tc3}.
We extended Vinen's equation to describe the diffusion and decay of the inhomogeneous vortex line density, $L(\boldsymbol{x},t)$:
\begin{equation}
\frac{dL(\boldsymbol{x},t)}{dt}=- \chi_2  \frac{ \kappa}{2 \pi}L(\boldsymbol{x},t)^2+D  \nabla^2L(\boldsymbol{x},t).          
\end{equation}
Here, $ \kappa$ is the quantized circulation, $\chi_2$ is the decay coefficient in the original Vinen equation, and $D$ is the introduced diffusion coefficient.
Our numerical work found $ \chi_2$ to be approximately 0.3 for a vortex tangle at zero temperature \cite{tsubota00}.   
A comparison between the numerical solutions of Eq. (1) and the numerical simulation of the diffusion of a vortex tangle showed $D$ is of the order of $ \kappa$. 
Hence, the diffusion was so small that the tangle remained localized in the neighborhood of the vibrating structure.

This work has paved the way for the studies of inhomogeneous quantum turbulence.  
Rockinson {\it et al.} studied theoretically and numerically the evolution of a cluster of quantum vortices in a free space \cite{Rickinson}. 
Nemirovskii addressed theoretically diffusion of inhomogeneous vortex tangle \cite{Nemi}.
Pomyalov studied theoretically and numerically the dynamics of turbulent plugs in a channel counterflow and revealed how a localized vortex tangle diffused \cite{Pomy}.
By visualization experiments using tracer particles, Tang {\it et al.} observed superdiffusion of quantized vortices  and the related scaling law \cite{Tang}, whose picture was confirmed numerically by Yui {\it et al}.\cite{Yui22}

\subsection{Kelvin wave cascade}
As described in Section 2, the Kelvin wave cascade is thought to play an important role in the energy transfer of a vortex tangle \cite{Vinen02,Vinen10}.
However, it was not trivial to determine whether such a cascade actually occurred along a single vortex line.
Thus, we performed a simple simulation of this issue at zero temperature \cite{Kelvin}. 
We prepared a single initially rectilinear vortex. 
Kelvin waves were continuously excited on the vortex by a sinusoidal force, whose frequency was resonant with the lowest Kelvin wave mode.
Damping at the highest wavenumber was allowed by the spatial resolution of the simulations.
The nonlinear coupling between different Kelvin wave modes led to a net flow of energy to higher wave numbers and the development of a simple spectrum of Kelvin waves that was insensitive to the strength and frequency of the exciting drive.
Thus, a picture of the Kelvin wave cascade was confirmed.
In this study, we presented the energy spectrum $E_k \sim k^{-1}$.

After our work, Kelvin wave cascade was actively studied from the perspective of wave turbulence \cite{Naza}.
There was an important controversy on the statistical behavior between Kozik-Svistunov \cite{KS04,KS08} and L'vov -Nazarenko \cite{LN07,Laurie,LN10,Boue}.
Kozik and Svistunov assumed that Kelvin wave interaction was local and derived the energe spectrum $E_k \sim k^{-7/5}$ \cite{KS04}.
However,  Laurie {\it et al.} found that the interaction was actually nonlocal \cite{Laurie} and obtained the correct spectrum $E_k \sim k^{-5/3}$ \cite{LN10,Boue}. 
These analyses were based on the VFM, while Kelvin wave cascade is now studied by the Gross-Pitaevskii model too \cite{Krstu, Muller}.

\subsection{Quantum turbulence created by an oscillating object}
\begin{figure}[h]%
\centering
\includegraphics[width=0.9\textwidth]{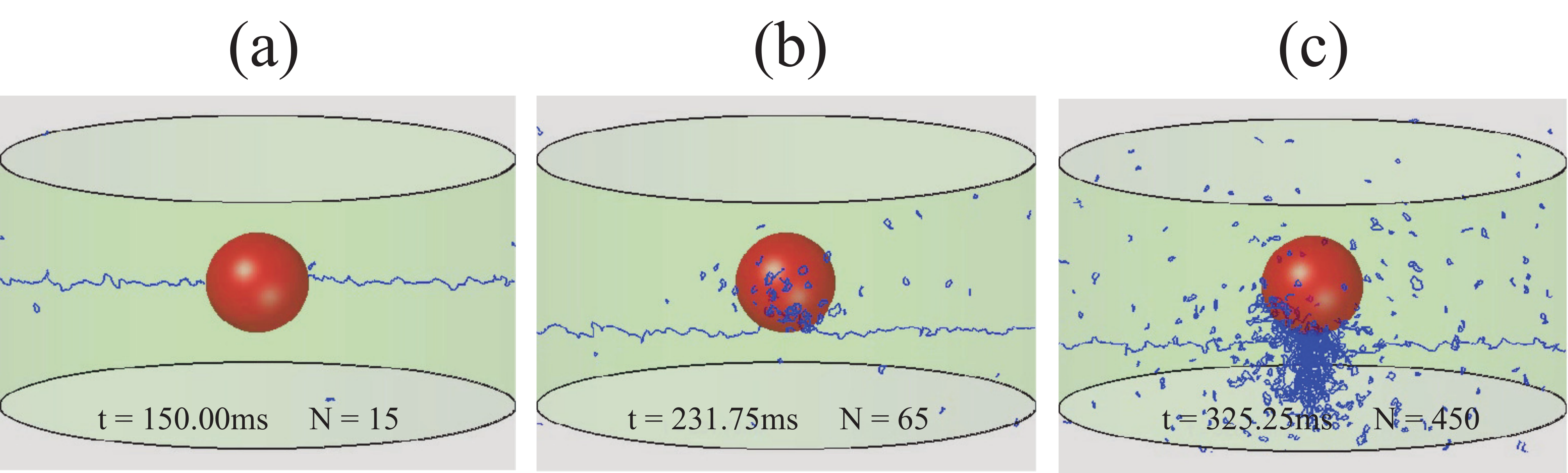}
\caption{Evolution of a vortex tangle around a sphere of radius ~100 $\mu$m in an oscillating superflow of 150 mm$^{-1}$ at ~200 Hz.
 [Reprinted figure with permission from R. H\"anninen, M. Tsubota, and W. F. Vinen, Phys. Rev. B {\bf 75}, 064502 (2007). Copyright (2007) by the American Physical Society.]}\label{Hanninen}
\end{figure}

Since the pioneering experiment that used an oscillating microsphere by J\"{a}ger  \cite{Jager}, quantum turbulence created by an oscillating object has become an important field \cite{VinenSkrbek}.  
The reason for this phenomenon is that a vortex tangle is created around an oscillating object when its velocity exceeds a critical value.
We numerically studied this at zero temperature, as shown in Fig. \ref{Hanninen} \cite{Hanninen}.
A stationary solid sphere was placed in a cylinder, and the remanant vortices were initially set between the sphere and both sides of the cylinder surface.
Subsequently, an oscillating superflow was applied in a direction perpendicular to the unperturbed remanent vortices.
Soon after the oscillation was established, Kelvin waves appeared on the remanent vortices (a). 
Continued oscillation led to increasing Kelvin wave amplitudes with increased nonlinear coupling to other wavenumbers. 
Then, at a sufficiently large Kelvin wave amplitude, reconnections occurred, resulting in the appearance of free vortex rings and vortex loops attached to the sphere (b).
In due course (c), a region of strong turbulence appeared on one or the other side of the sphere in the direction of the flow, and the turbulence apparently resulted from loops being pulled out from the surface of the sphere.
This finding has served as a guideline for subsequent research on this topic.

\subsection{Coupled dynamics of tracer particles and quantized vortices}
\begin{figure}[h]%
\centering

\includegraphics[width=0.5\textwidth]{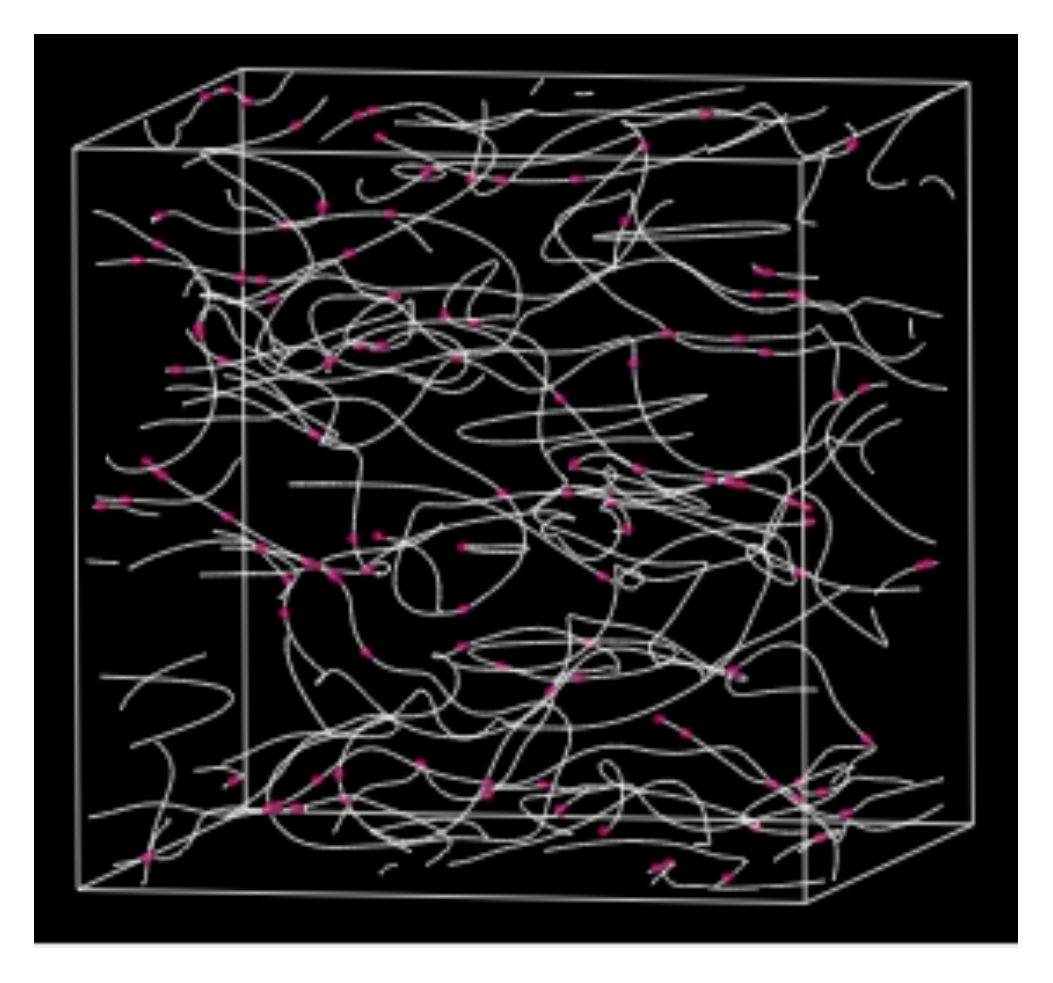}
\caption{Snapshot of the coupled dynamics of the tracer particles and the vortices at $T$ = 1.9 K.
All of the 120 particles (red points) were trapped by vortices. The direction of the normal flow, $v_n$ = 0.30 cm/s, was upward and that of superflow, $v_s$, was downward. 
The vortex line density was approximately 6000 cm$^{-2}$.
 [Reprinted figure with permission from Y.  Mineda, M. Tsubota,  Y.  A. Sergeev, C. F. Barenghi and W. F. Vinen, Phys. Rev. B {\bf 87}, 174508 (2013). Copyright (2013) by the American Physical Society.]} \label{Mineda}
\end{figure}

The visualization experiments used fine solid particles as tracers  \cite{WeiGuo}.
Paolett {\it et al.} succeeded in visualizing the characteristic motion of fine particles in thermal counterflow \cite{Paoletti}. 
The tracer particles were divided into two groups: those that moved freely in the direction of the normal fluid and those that were trapped into vortices and moved in the opposite direction. 
The velocity distribution of the trapped particles was broader, reflecting an irregular motion of the vortices.
To account for these observations, we formulated the coupled dynamics of particles and quantized vortices and performed numerical simulations \cite{Mineda}. 
The numerical simulations showed the characteristic coupled motion of the particles and vortices (Fig. \ref{Mineda}), and agreed reasonably well with observations of the velocity distributions of the tracer particles.

\section{Summary}
Remembering my work with Vinen, I reaffirm his huge contribution to the field of quantum turbulence and recall vivid scenes of discussions on each topic with him.
These works were not only important at that time, but also directed subsequent developments.
I should say that they are evidence and tributes to the far-sightedness of Vinen.



\begin{thebibliography}{}
\bibitem{hallvinen1} H. E. Hall, W. F. Vinen, Proc. Roy. Soc. A {\bf 238}, 204 (1956) 
\bibitem{hallvinen2} H. E. Hall, W. F. Vinen, Proc. Roy. Soc. A {\bf 238}, 215 (1956) 
\bibitem{tc1} W. F. Vinen, Proc. Roy. Soc. A {\bf 240}, 114 (1957) 
\bibitem{tc2} W. F. Vinen, Proc. Roy. Soc. A {\bf 240}, 128 (1957) 
\bibitem{tc3} W. F. Vinen, Proc. Roy. Soc. A {\bf 242}, 493 (1957) 
\bibitem{tc4} W. F. Vinen, Proc. Roy. Soc. A {\bf 243}, 400 (1957)
\bibitem{circulation} W. F. Vinen, Proc. Roy. Soc. A {\bf 260}, 218 (1961)  
\bibitem{davis} S. L. Davis, P. C. Hendry, and P. V. E. McClintock, Physica B {\bf 280}, 43 (2000)
\bibitem{Vinen02} W. F. Vinen, J. J. Niemela, J. Low Temp. Phys. {\bf 128}, 167 (2002)
\bibitem{Vinen10} W. F. Vinen, J. Low Temp. Phys. {\bf 161}, 419 (2010)
\bibitem{JLTPreview} M. Tsubota, K. Fujimoto and S. Yui, J. low Temp. Phys. {\bf 188}, 119 (2017)
\bibitem{Maurer98} J. Maurer, P. Tabeling, Europhys. Lett.  {\bf 43}, 29 (1998)
\bibitem{Stalp99} S. R. Stalp, L. Skrbek, R. J. Donnelly, Phys. Rev. Lett. {\bf 82}, 4831 (1999)
\bibitem{Frisch} U. Frisch, {\it Turbulence} (Cambridge University Press, Cambridge,1995)
\bibitem{davidson} P. A. Davidson, {\it Turbulence: An Introduction for Scientists and Engineers}, 2nd edn. (Oxford University Press, Oxford 2015)
\bibitem{Vinen00} W. F. Vinen, Phys. Rev. B {\bf 61}, 1410 (2000)
\bibitem{tsubota00} M. Tsubota, T. Araki,  S. K.  Nemirovskii,  Phys. Rev. B {\bf 62}, 11751 (2000)
\bibitem{Bagnato} M. C. Tsatsos, P. E. S. Tavares, A. Cidrim, A. R. Fritsch, M. A. Caracanhas, F. E. A. dos Santos, C. F. Barenghi, V. S. Bagnato, Phys. Rep. {\bf 622}. 1 (2016)
\bibitem{Walmsley07} P. M. Walmsley, A. I. Golov, H. E. Hall, A. A. Levchenko and W. F. Vinen, Phys. Rev. Lett. {\bf 99}, 265302 (2007)
\bibitem{Walmsley08} P. M. Walmsley and  A. I. Golov, Phys. Rev. Lett. {\bf 100}, 245301 (2008)
\bibitem{GolovPNAS} P. M. Walmsley, D. Zmeev, F. Pakpour and  A. I. Golov, Proc. Natl. Acad. Sci. U.S.A. {\bf 111}, 4691 (2014)
\bibitem{Araki02} T. Araki, M. Tsubota, S. K.  Nemirovskii,  Phys. Rev. Lett. {\bf 89}, 145301 (2002)
\bibitem{Nore} C. Nore, M. Abid, M. E. Brachet, Phys. Rev. Lett. {\bf 78}, 3896 (1997); Phys. Fluids \textbf{9}, 2644 (1997)
\bibitem{tough} J. T. Tough, in {\it Prog. Low Temp. Phys.}, Vol. 8, ed. By C. J. Gorter (North-Holland, Amsterdam, 1982), pp. 133-219
\bibitem{Melotte} D. J. Melotte and C. F. Barenghi, Phys. Rev. Lett. {\bf 80}, 4181 (1998)
\bibitem{Adachi} H. Adachi, S. Fujiyama and M. Tsubota, Phys. Rev. B {\bf 81}, 104511 (2010)
\bibitem{Schwarz88} K. W. Schwarz, Phys. Rev. B {\bf 38}, 2398 (1988)
\bibitem{WeiGuo} W. Guo, M. La Manta, D. P. Lathrop and S. W. Van Sciver,  Proc. Natl. Acad. Sci. U.S.A. {\bf 111}, 4653 (2014)
\bibitem{Bewley} G. P. Bewley, D. P. Lathrop and K. R. Sreenivasan, Nature {\bf 441}, 588 (2006)
\bibitem{Guo10} W. Guo, S. B. Cahn, J. A. Nikkel, W. F. Vinen and D. N. McKinsey, Phys. Rev. Lett. {\bf 105},045301 (2010)
\bibitem{Marakov} A. Marakov, J. Gao, W. Guo, S. W. Van Sciver, G. G. Ihas, D. N. McKinsey, and W. F. Vinen, Phys. Rev. B {\bf 91},094503 (2015)
\bibitem{KIvotides00} D. Kivotides, C. F. Barenghi and D. C. Samuels, Science {\bf 290}, 777 (2000)
\bibitem{KIvotides07} D. Kivotides, Phys. Rev. B {\bf 76}, 054503 (2007)
\bibitem{Yui18} S. Yui, M. Tsubota and H. Kobayashi, Phys. Rev. Lett. {\bf 120}, 155301 (2018)
\bibitem{Mastracci18} B. Mastracci and W. Guo, Phys. Rev. Fluids {\bf 3}, 063304 (2018)
\bibitem{Mastracci19} B. Mastracci and W. Guo, Phys. Rev. Fluids {\bf 4}, 023301 (2019)
\bibitem{Mastracci19-2} B. Mastracci, S. Bao, W. Guo and W. F. Vinen, Phys. Rev. Fluids {\bf 4}, 083305 (2019)
\bibitem{Tang20} Y. Tang, S. Bao, T. Kanai and W. Guo, Phys. Rev. Fluids {\bf 5}, 84602 (2020)
\bibitem{Yui20} S. Yui, H. Kobayashi, M. Tsubota and W. Guo, Phys. Rev. Lett. {\bf 124}, 155301 (2020)
\bibitem{VinenSkrbek} W. F. Vinen and L. Skrbek,  Proc. Natl. Acad. Sci. U.S.A. {\bf 111}, 4699 (2014)
\bibitem{Diffusion} M. Tsubota, T. Araki  and W. F. Vinen,  Physica B {\bf 329-333}, 224 (2003)
\bibitem{Rickinson} E. Rickinson, N. G. Parker, A. W. Baggaley and C. F. Barenghi, Phys. Rev. A {\bf 98}, 023608 (2018)
\bibitem{Nemi} S. K. Nemirovskii, Phys. Rev. B {\bf 81}, 064512 (2010)
\bibitem{Pomy} A. Pomyalov, Phys. Rev. B {\bf 101}, 134515 (2020)
\bibitem{Tang} Y. Tang, S. Bao and W. Guo, Proc. Natl. Acad. Sci. U.S.A. {\bf 118}, e2021957118 (2021)
\bibitem{Yui22} S. Yui, Y. Tang, W. Guo, H. Kobayashi and M. Tsubota, Phys. Rev. Lett. {\bf 129}, 025301 (2022)
\bibitem{Kelvin} W. F. Vinen, M. Tsubota and A. Mitani, Phys. Rev. Lett. {\bf 91}, 135301 (2003)
\bibitem{Naza} S. Nazarenko, {\it Wave Turbulence}, Lecture Notes in Physics, Vol. 825 (Springer, Heidelberg, 2011)
\bibitem{KS04} E. Kozik and B. Svistunov, Phys. Rev. Lett. {\bf 92}, 035301 (2004)
\bibitem{KS08} E. Kozik and B. Svistunov, Phys. Rev. B {\bf 77}, 060502(R) (2008)
\bibitem{LN07} V. S. L’vov, S. V. Nazarenko, and O. Rudenko, Phys. Rev. B {\bf 76}, 024520 (2007)
\bibitem{Laurie} J. Laurie, V. S. L’vov, S. Nazarenko, and O. Rudenko, Phys. Rev. B {\bf 81}, 104526 (2010)
\bibitem{LN10} V. S. L’vov and S. Nazarenko,  JETP Lett. {\bf 91}, 428 (2010)
\bibitem{Boue} L. Bou\'{e}, R. Dasgupta, J. Laurie, V. S.  L’vov, S. Nazarenko and I. Procaccia,  Phys. Rev. B {\bf 84}, 064516 (2011)
\bibitem{Krstu} G. Krstulovic, Phys. Rev. B {\bf 86}, 055301(R) (2012)
\bibitem{Muller} N. P. M\"{u}ller and G. Krstulovic, Phys. Rev. B {\bf 102}, 134513 (2020)
\bibitem{Jager} J. J\"{a}ger, B. Schuderer and W. Schoepe, Phys. Rev. Lett. {\bf 74}, 566 (1995)
\bibitem{Hanninen} R. H\"anninen, M. Tsubota, and W. F. Vinen, Phys. Rev. B {\bf 75}, 064502 (2007)
\bibitem{Paoletti} M. S. Paoletti, R. B. Fiorito, K. R. Sreenivasan, and D. P. Lathrop, J. Phys. Soc. Jpn. {\bf 77}, 111007 (2008)
\bibitem{Mineda} Y.  Mineda, M. Tsubota,  Y.  A. Sergeev, C. F. Barenghi and W. F. Vinen, Phys. Rev. B {\bf 87}, 174508 (2013)
\end{thebibliography}
\end{document}